# Development of Silicon Strip Detectors for a Medium Energy Gamma-ray Telescope

P. Bloser[a,*], F. Schopper[a], R. Andritschke[a], G. Kanbach[a], A. Zoglauer[a], P. Lechner[b]

[a]*Max-Planck-Institut für extraterrestrische Physik, Giessenbachstrasse, 85748 Garching, Germany*

[b]*PNSensor GmbH, Römerstr. 28, 80803 München, Germany*

**Abstract**

We report on the design, production, and testing of advanced double-sided silicon strip detectors under development at the Max-Planck-Institute as part of the Medium Energy Gamma-ray Astronomy (MEGA) project. The detectors are designed to form a stack, the "tracker," with the goal of recording the paths of energetic electrons produced by Compton-scatter and pair-production interactions. Each layer of the tracker is composed of a 3 × 3 array of 500 μm thick silicon wafers, each 6 cm × 6 cm and fitted with 128 orthogonal p and n strips on opposite sides (470 μm pitch). The strips are biased using the punch-through principle and AC-coupled via metal strips separated from the strip implant by an insulating oxide/nitride layer. The strips from adjacent wafers in the 3 × 3 array are wire-bonded in series and read out by 128-channel TA1.1 ASICs, creating a total 19 cm × 19 cm position-sensitive area. At 20°C a typical energy resolution of 15-20 keV FWHM, a position resolution of 290 μm, and a time resolution of ~1 μs is observed.

**1. Introduction**

In the field of gamma-ray astronomy there is an urgent need for a new mission sensitive from 0.4-50 MeV to follow up on the success of the COMPTEL telescope [1] on the recently-deorbited Compton Gamma-Ray Observatory satellite. This so-called medium-energy gamma-ray regime is crucial for the study of a rich variety of high-energy astrophysical processes. The goal of the Medium Energy Gamma-ray Astronomy (MEGA) project is to meet this need, using modern detector technology to improve on the sensitivity of COMPTEL by a factor of ten.

---

[*] Corresponding author. Tel.: +49-89-30000-3854; fax: +49-89-30000-3606; e-mail: bloser@mpe.mpg.de.

Two physical processes dominate the interaction of photons with matter in the medium-energy gamma-ray band: Compton scattering at low energies, and electron-positron pair production at high energies, with the changeover at 5-10 MeV for most detector materials. In both cases the primary interaction produces long-range secondaries whose directions and energies must be determined in order to reconstruct the incident photon. MEGA, like previous Compton and pair telescopes, will employ two separate detectors to accomplish this task: a *tracker*, in which the initial Compton scatter or pair conversion takes place, and a *calorimeter*, which absorbs and measures the energy of the secondaries (see Figure 1). In the case of Compton interactions, the incident photon scatters off an electron in the tracker, which measures the interaction position and the energy imparted to the electron. The scattered photon interaction point and energy are recorded in the calorimeter; from the positions and energies of the two interactions the incident photon angle $\varphi$ may be computed from the Compton equation. The primary photon's incident direction is then constrained to an "event circle" on the sky. For incident energies above about 2 MeV the recoil electron usually receives enough energy to penetrate several layers, allowing it to be tracked. This further contrains the incident direction of the photon ("reduced event circle" in Figure 1). In the case of pair production, the incident photon converts into an electron-positron pair in the tracker. These two particles are tracked so as to determine the incident photon direction, and then absorbed in the tracker and/or the calorimeter to measure the total energy.

While the job of the calorimeter is straightforward and will be adequately performed by scintillators [2], the tracker must accomplish several different tasks: 1) act as the scattering medium for Compton interactions; 2) measure the scattering location and the energy imparted to the recoil electron; 3) act as the conversion medium for pair production; 4) provide a large interaction volume for both processes to achieve good sensitivity; 5) record the tracks and energy deposits of all secondary particles, both electron-positron pairs and Compton recoil electrons; 6) provide a fast timing signal to be used in a coincident trigger with the calorimeter; and 7) operate without elaborate cooling and with reasonable power consumption. Thus good position, energy, and time resolution are required simultaneously in a large volume.

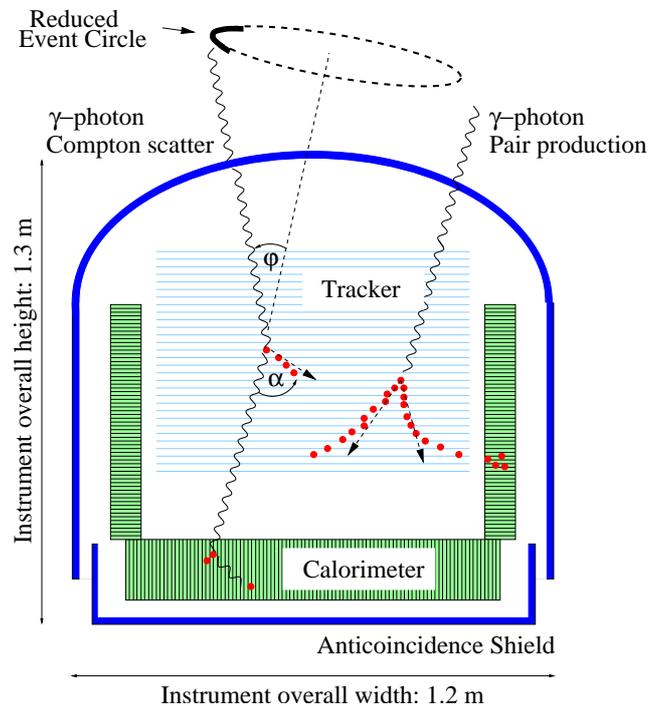

Fig. 1. Measurement principles of the MEGA telescope.

The logical choice of detector technology for the tracker is a stack of double-sided silicon strip detectors. COMPTEL relied on large liquid scintillator blocks read out by photomultiplier tubes for its scattering detectors, which limited the energy resolution to ~45 keV FWHM and the spatial resolution to ~5.4 cm FWHM [1]. The availability of low-noise, low-power multichannel preamplifier chips makes it feasible to read out a large number of channels within the space and power constraints of a space mission, so that the scintillator technology of COMPTEL may be replaced with semiconductors. The use of strips allows large areas to be covered with a reasonable number of channels. Although silicon strip detectors have been used as particle trackers in accelerator experiments for some time, the requirements of the MEGA tracker impose special demands, namely two-dimensional position information from each layer, good energy resolution, and low power consumption. We have constructed a small prototype of the MEGA experiment at the Max-Planck-Institute for Extraterrestrial Physics (MPE) to demonstrate the technology required for the full satellite mission.  We report here on the design and performance of the silicon strip detectors used in the tracker prototype.

Table 1  Effect of Doppler broadening on Compton scatter angle distribution from GEANT3 simulations

| Energy of the Primary photon (keV) | 140 | 511 | 1000 | 2000 | 5000 |
|---|---|---|---|---|---|
| FWHM of the Doppler-broadened scatter angle distribution (degrees) for untracked/tracked events | 2.1 / - | 0.65 / - | 0.3 / 0.5 | 0.15 / 0.25 | 0.1 / 0.15 |
| Equivalent FWHM energy resolution (keV) for untracked/tracked events | 1.5 / - | 2.1 / - | 2.4 / 4.5 | 3.5 / 6.5 | 3.7 / 9.0 |

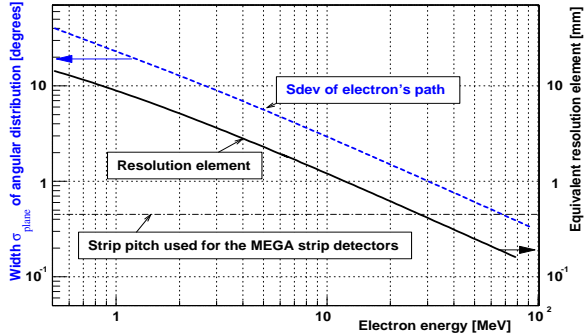

Fig. 2. Standard deviation of electron´s path due to Molière scattering after traversing 250 μm of silicon (left axis, dotted curve), and resolution element giving same standard deviation at 10 mm distance (right axis, solid curve).  The strip pitch of the MEGA detectors is also indicated.

## 2. Requirements for the tracker

The required position, energy, and time resolution of the tracker elements are determined by the relative demands of pair and Compton imaging.

### 2.1. Position resolution

The tracker position resolution is constrained by the requirements of pair imaging.  The angular resolution of the incident photon is limited by the precision with which the electron and positron tracks can be reconstructed.  The track reconstruction below 100 MeV is in turn limited by the small-angle scattering of the charged particles in the detector material.  For a silicon wafer thickness of 500 μm, a typical electron will traverse 250 μm before exiting the detector.  The standard deviation of the electron´s path due to Molière scattering [3] in this case is shown in Figure 2.  Also shown is the width $W$ of the resolution element on the adjacent layer that would record the same standard deviation in electron direction in one dimension $(\sigma = W/\sqrt{12}$, where no interpolation between strips is used) for a 10 mm separation between layers and initially normal tracks.  ($W$ is divided again by $\sqrt{2}$ since we make a measurement on each layer.)  The actual pitch need not be any finer than this.  From Figure 2, a pitch of 0.5 mm is sufficient for photon energies below 50 MeV.  Note that we have made the usual assumption that the electron carries half the photon energy.

Whereas the resolution element pitch is set by pair tracks, which traverse several layers, Compton scatter interactions take place in a single layer, and at low energies they do not produce tracks.  Therefore for Compton imaging each detector must provide both x and y positional information.

### 2.2. Energy resolution

The required tracker energy resolution is set by Compton imaging considerations.  The error in the reconstructed incident photon direction depends on the measurement error in both the energy imparted to the scattering electron and the energy of the scattered photon.  There is a fundamental limit to the accuracy with which the scatter angle can be derived from the measured energies, due to the fact that the electron is not initially at rest.  Rather, the unknown momentum of the electron within its atomic energy shell leads to a "Doppler broadening" of the relative energies of the electron and scattered photon, with a subsequent broadening in the reconstructed incident photon direction.  Table 1 shows the FWHM of the Doppler-broadened scatter angle distribution, as given by GEANT3 simulations using the GLECS[1] package to include the electron momentum. Also shown is the equivalent energy resolution in the scattering detector that would produce the same error. The tracker energy resolution need not be better than this. Both tracked and non-tracked events are shown, as the latter typically have larger scatter angles and larger errors. Note also that when the recoil electron produces a track, the relevant quantity is the sum of the energies in the individual layers, with the corresponding combination of their errors. The numbers indicate that in the MeV energy range there is no great merit in going much beyond the energy resolution achievable with silicon strip detectors, 10-15 keV FWHM. We note also that the effects of Doppler broadening grow worse with increasing atomic number, reducing the advantage of the high energy resolution of materials like germanium.

*2.3. Time Resolution*

The COMPTEL telescope relied on the relative timing of hits in the scattering detector and the calorimeter (the "time of flight") to reduce background, which, for a detector separation of 1.5 m, mandated nanosecond time resolution. To improve sensitivity, the tracker and calorimeter in MEGA will be much closer together, making such time-of-flight measurements impossible. The timing requirements of the tracker are therefore more modest, although a short coincidence window reduces the likelihood of chance coincidences. The charge collection time in thin semiconductors (50-100 ns) is sufficient if the readout circuit includes a fast-shaped channel for triggering.

**3. Detector Design**

Double-sided silicon strip detectors are the sensible choice to fulfill the above requirements for MEGA. The strip detectors for the prototype tracker were designed by the Max-Planck-Institute Semiconductor Lab and produced by EURISYS. The

---

GEANT Low-Energy Compton Scattering package; http://www.batse.msfc.nasa.gov/actsim/glecs.html

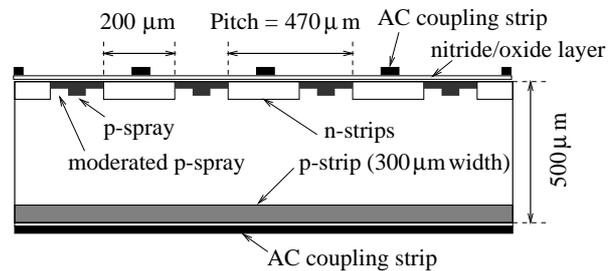

Fig. 3. Cross section of the MEGA strip detector geometry.

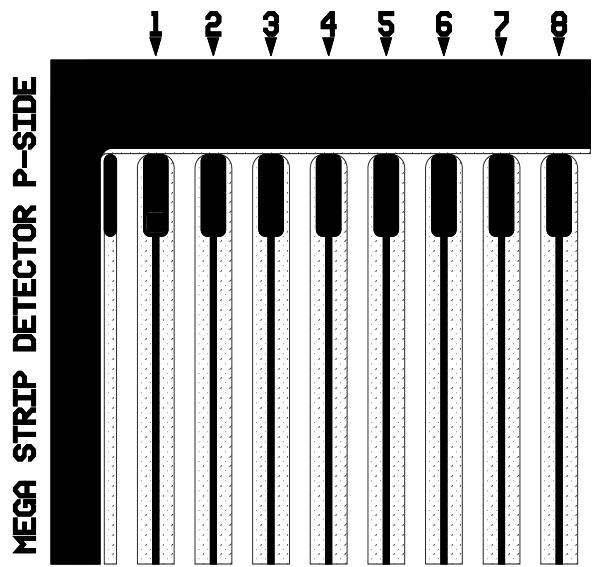

Fig 4. Corner of the p-side layout, showing bias ring, guard rings, strips, and AC coupling strips.

cross-sectional layout of the detectors is shown in Figure 3. The detector wafers were produced from n-type substrate required to have a resistivity greater than 5 k$\Omega$-cm (measured values were typically 10 $\Omega$-cm). Each individual wafer is 6 cm × 6 cm, 500 $\mu$m thick, and fitted with 128 orthogonal p and n strips on opposite sides. The strip pitch is 470 $\mu$m, with a strip width of 300 $\mu$m on the p-side and 200 $\mu$m on the n-side. The n-side strips are separated by a non-masked floating p-implantation, or "p-spray." The p-spray concentration is lowered along the edges of the strips to reduce the electric field and strip capacitance, a technique known as "moderated p-spray."

The strips are biased via the punch-through principle. The punch-through contacts are formed across a gap between the ends of each strip and a common bias ring. The gap width is 8 μm on the p-side and 12 μm on the n-side. The p-side bias ring is surrounded by ten narrow guard rings to bring the bias voltage applied to the bias ring down to zero at the cut crystal edge. The strips are AC-coupled on the wafer itself by metal strips 50 μm wide, separated from the implant by insulating layers of silicon oxide (1500 Å) and silicon nitride (900 Å). For diagnostic reasons every eighth strip has an additional ohmic contact. Figure 4 shows the layout of the p-side, including bias ring, guard rings, strips, and AC coupling strips.

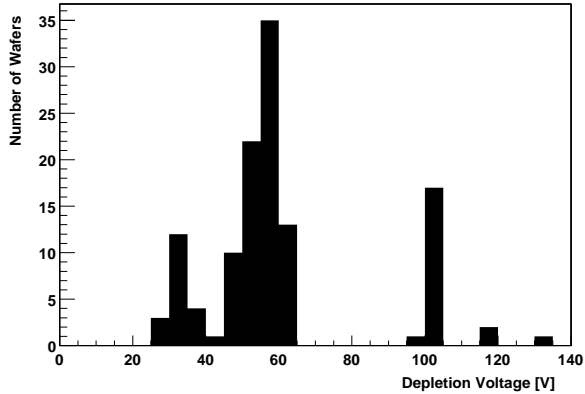

Fig. 5. Distribution of depletion bias voltages. Most wafers were only measured up to 100 V, so that the depletion voltages of the 16 wafers plotted at exactly 100 V is unkown.

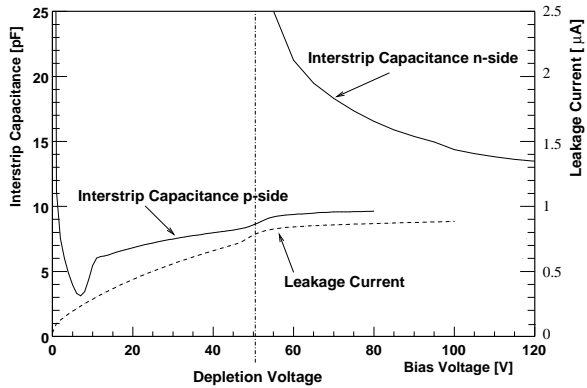

Fig. 6. Measured interstrip capacitance for p- and n-sides as a function of bias voltage.

## 4. Individual Wafer Results

Measurements were performed on the individual detector wafers with a probe station at the Semiconductor Lab. In total 171 wafers were delivered to the Lab by EURISYS. Of these, 38 (22%) were found to have more than two AC coupling strips shorted to the implant strip due to holes or impurities in the oxide/nitride layer. These wafers were discarded. In addition, 25 wafers (14%) had a leakage current higher than 3.5 μA at 60 V bias voltage and ambient room temperature (~25 °C). These were also discarded. The remaining wafers were characterized as follows:

### 4.1. Depletion Bias Voltage

In order for the strips on the n-side to be electrically separated, the wafer must be fully depleted. The bias voltage resulting in full depletion was found to vary greatly from wafer to wafer. Figure 5 shows the distribution in depletion voltages; the wafers fall into three groups from roughly 25-40 V, 45-65 V, and 100-140 V. In general, wafers from the same production batch had similar depletion voltages, indicating that the differences stem from variations in the starting materials´ resistivity. Since several wafers are mounted together in each tracker layer, and each layer can receive only one bias voltage, the differences must be carefully taken into account. This is especially true since some wafers break down when operated more than 10-20 V over their depletion voltage, giving only a narrow operating range. For the prototype tracker the layers are biased at either 55 V, 65 V, or 145 V, depending on the wafers they contain.

### 4.2. Interstrip Capacitance

The capacitance of a strip detector is dominated by the capacitance between adjacent strips. This interstrip capacitance was measured as a function of bias voltage for the p- and n-sides (Figure 6). The p-side capacitance above the depletion voltage is independent of the bias, and the measured value of ~9 pF is in good agreement with device simulations. Simulations also predict a constant value close to 9 pF for the n-side, but this is not observed. Rather, the n-side interstrip capacitance lies at ~20 pF slightly above the depletion voltage and decreases slowly with increasing bias. This may indicate an increasing width of the lateral depletion between n-strips and the p-spray region, resulting in a decrease of the interstrip capacitance. The reason for the overall higher capacitance on the n-side is unclear.

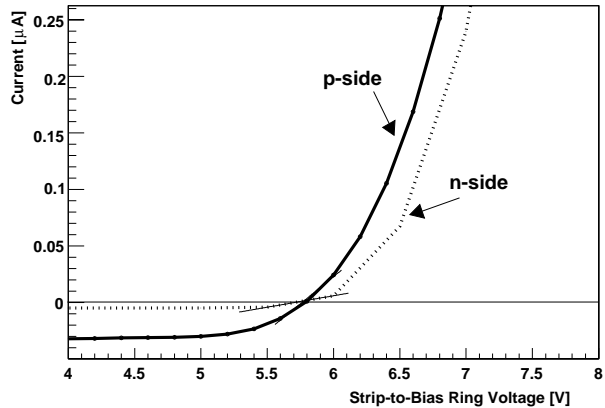

Fig. 7. I-V curves of the punch-through contacts on the p-side (solid) and n-side (dotted). The slope at zero-crossing indicates the effective bias resistance.

Device simulations predict a capacitance between strips and backside of ~5.5 pF.

### 4.3. Punch-Through Biasing

The dynamic resistance of the punch-through bias contact was measured by applying a voltage difference between one of the DC-coupled strips and the bias ring (while the detector was depleted) and measuring the resulting current. The effective bias resistance is determined from the slope of the current-voltage (I-V) curve at the point when zero current flows in the current meter, since this imitates the actual operating condition when the meter is not present. Typical measured values were 11 MΩ on the p-side and 50 MΩ on the n-side, as shown in Figure 7. These bias resistance values are high enough to be unimportant for noise considerations, assuming the punch-through contact does indeed behave like a resistor (see Section 5.1).

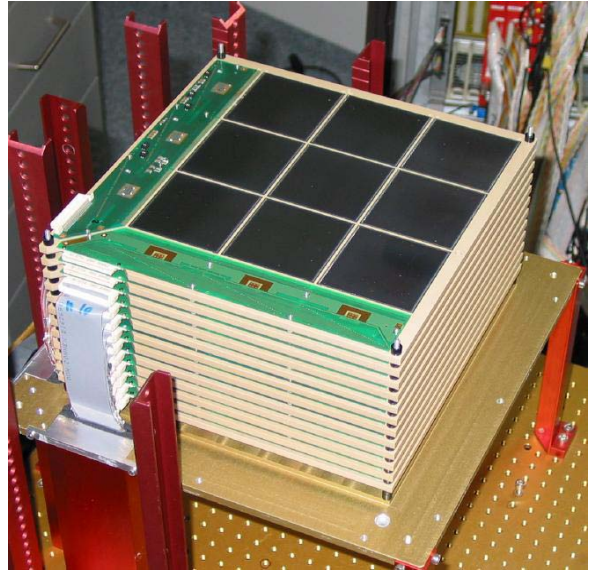

Fig. 8. Assembled prototype tracker. Visible in the top layer are the 9 wafers and p-side readout chips, mounted on the hybrid board at left. The n-side chips, reading out the strips on the bottom surface, are mounted on the underside of the board at the front.

### 5. Tracker Layer Results

Each prototype tracker layer is composed of a 3 × 3 array of wafers glued onto a plastic PEEK frame. This gives a nearly continuous sensitive surface area 18.9 cm across (~357 cm$^2$) with a 1.5 mm gap between wafers. The strips of adjacent wafers are wire-bonded in series so that each layer effectively consists of 384 strips, 18.9 cm long, on both the top and bottom surfaces. Each ladder of three wafers is read out by a 128-channel TA1.1 ASIC chip manufactured by IDE AS Corp., Norway. The TA1.1 is a low-noise, low-power charge-sensitive preamplifier and shaper circuit with sample-and-hold and multiplexed analog readout. It is self-triggering, with a simple threshold discriminator and loadable bitmask to block individual channels. The measured power consumption is ~0.5 mW/channel using the control voltages appropriate for MEGA. Three chips are required to read out the p- and n-sides, respectively; these are mounted on hybrid boards at the edge of each layer. Figure 8 shows the assembled prototpye tracker. The following results were

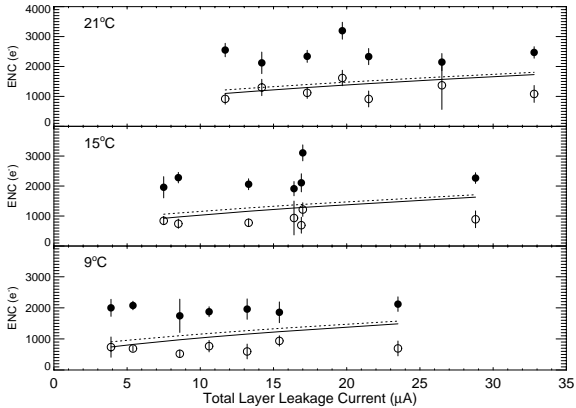

Fig. 9. Measured noise for seven layers at three temperatures, plotted versus leakage current and compared with calculated values. The p-side (open symbols) is in good agreement with expectations (solid line), while the n-side (closed symbols) is a factor of two above the expected values (dotted line). Error bars indicate the standard deviation of the noise values within each layer.

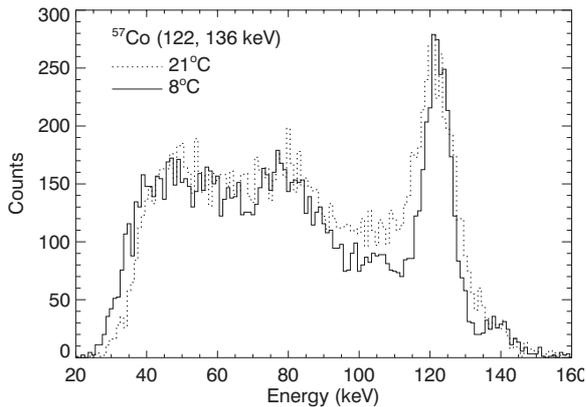

Fig. 10. $^{57}$Co spectra measured by one layer (p-side) at 21°C and 8°C. The photopeak FWHM values are 15.7 keV and 10.3 keV, respectively.

obtained for the seven working layers at the time of this writing:

### 5.1. Noise and Energy Resolution

The most important detector parameter for the MEGA tracker is the readout noise, as this determines the energy resolution and, therefore, the angular resolution of the telescope for Compton-scatter events. The expected noise, expressed as equivalent noise charge (ENC) in electrons, may be calculated as follows [4]:

$$ENC^2 = \left(ENC_{base} + ENC_{slope} \times C_{tot}\right)^2 + 2A\tau\left(\frac{I_{leak}}{q} + \frac{kT}{qR_{bias}}\right)$$

Here $ENC_{base}$ is the baseline electronic noise independent of capacitance, while $ENC_{slope}$ describes the noise proportional to the total capacitance $C_{tot}$. The parameter $A$ is a filter coefficient and is equal to 1.3 for semigaussian shaping with time constant $\tau$. For the TA1.1 chip, $ENC_{base}$ = 165 e$^-$ and $ENC_{slope}$ = 7 e$^-$/pF at a shaping time $\tau$ = 2 µs. The last two terms describe the noise from the leakage current $I_{leak}$ and the Johnson noise in the bias resistance $R_{bias}$, where $q$ is the electron charge, $k$ is the Boltzmann constant, and $T$ is the temperature.

The noise for all seven working tracker layers was recorded at three different temperatures (21°C, 15°C, and 9°C). In Figure 9 we compare the average measured noise values from the p- and n-sides of each layer to those expected from the above equation as a function of leakage current and temperature. The total capacitance $C_{tot}$ was taken to be the measured p- or n-side interstrip capacitance per wafer plus the simulated strip-to-backside capacitance (Section 4.2), multiplied by three for the three concatenated wafers in each strip. The measured p- or n-side bias resistance per wafer $R_{bias}$ was used (Section 4.3), divided by three for the three parallel contributions from each wafer. The measured leakage current for each wafer was assumed to be distributed evenly over all strips. Since it is not possible to measure what fraction of the leakage current flows over the guard rings, this represents an upper limit to the current in each strip, and so the calculated noise will also be an upper limit. While the p-side noise values are in good agreement with or slightly lower than the calculated values, the n-side noise is a factor of ~2 too high. The cause of this discrepancy is unknown; the higher interstrip capacitance of the n-side has only a small effect on the predicted noise (dotted line in Figure 9). One possibility is that the punch-through bias contacts do not in fact act like resistors but rather like FETs, which behave differently on the n-side than on the p-side and result in higher noise.

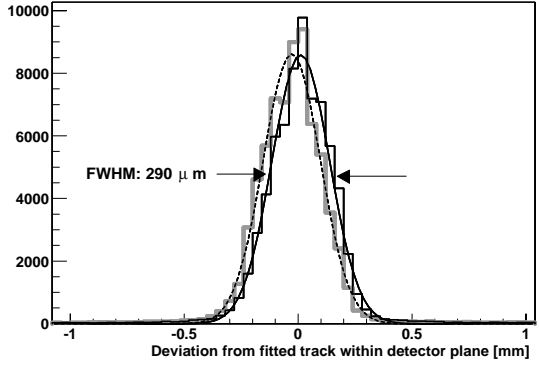

Fig. 11. Tracker position resolution measured with muon tracks. Plotted is the deviation of the measured hit position in one layer from the fitted track for the n-side (shaded histogram, dotted curve) and p-side (solid histogram and curve).

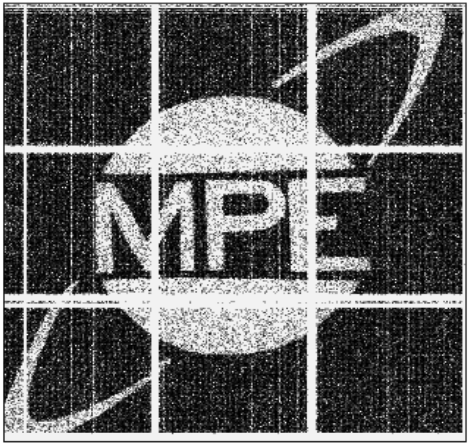

Fig. 12. Shadow mask image taken with a lead mask and $^{57}$Co source.

Table 2  $^{57}$Co (122 keV) photopeak widths

| Layer | FWHM @ 20°C (keV) p-side | FWHM @ 20°C (keV) n-side | FWHM @ 8°C (keV) p-side | FWHM @ 8°C (keV) n-side |
|---|---|---|---|---|
| 1 | 15.8 | 29.4 | 10.3 | 24.1 |
| 2 | 17.8 | 28.1 | 12.9 | 23.2 |
| 3 | 18.8 | 35.7 | 17.7 | 29.1 |
| 4 | 26.6 | 32.6 | 23.4 | 25.0 |
| 5 | 28.3 | 42.2 | 17.3 | 28.6 |
| 6 | 26.4 | 35.2 | 20.8 | 29.8 |
| 7 | 21.4 | 30.4 | 14.6 | 29.6 |

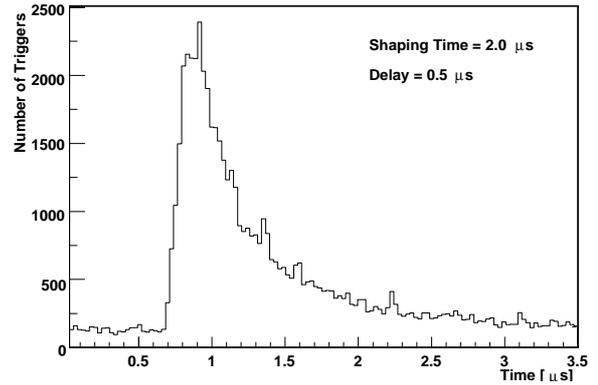

Fig. 13. Time delay of tracker triggers relative to a fast plastic scintillator.

Figure 10 shows $^{57}$Co energy spectra (122 and 136 keV lines) recorded by one layer at 21°C and 8°C. The measured 122 keV photopeak FWHM values are 15.7 keV and 10.3 keV, respectively. Only energy values from the p-side were used. Signals from neighboring strips were combined in producing these spectra, but no other cuts were applied to the data. The average 122 keV photopeak FWHM values over all strips for each layer are shown in Table 2. On the p-sides at least, most layers have photopeak FWHM values between 15-25 keV at room temperature, and 10-20 keV with moderate cooling. This energy resolution approaches the practical limit for Compton imaging set by Doppler broadening in the MeV range in silicon (Section 2.2).

*5.2. Position Resolution*

The position resolution of the tracker was measured using muon tracks. With the seven layers stacked in the tracker configuration (Figure 8), events were recorded using a coincident trigger between the top and bottom layer. A straight line was fitted to tracks containing hits in at least five layers, and the difference between the recorded hit position and the fitted track intersection (taken as the "true" position) in a given layer was measured. For hits containing multiple adjacent strips a simple weighted average was used for the position. The distribution of deviations from the "true" position for a typical layer is shown in Figure 11. The FWHM of the distribution is 290 μm for both p- and n-sides, slightly better than the $2.35 \times (470\ \mu m / \sqrt{12}) = 319$ μm expected for the MEGA strip pitch for single-strip hits. Note that this width includes the error in the fitted track direction caused by the position resolution of the other layers, so that the true resolution is probably somewhat better than this. The measured position resolution is sufficient given the limits set by Molière scattering of particles (Section 2.1). Figure 12 shows a shadow mask image taken by one layer with 122 keV photons, demonstrating the imaging capability of the MEGA strip detectors in the hard X-ray range.

*5.3. Time Resolution*

The time resolution of the tracker layers was measured relative to a plastic scintillator with constant fraction timing using a positron emitter ($^{22}$Na) placed between the two detectors. Figure 13 shows the time distribution of trigger signals in the tracker, which is dominated by time walk due to the simple level trigger of the TA1.1 chip. While adequate for lab tests, the timing spread of ~1 µs is rather long for low-background coincidence measurements. Future versions of the TA chips will include a fast (100 ns) shaper for each channel for triggering purposes.

## 6. Conclusions

The silicon strip detectors produced for the MEGA prototype have demonstrated energy, position, and time resolution appropriate for a next-generation Compton and pair telescope. They will now be combined with the calorimeter prototype for tests of the complete MEGA telescope concept.

**Acknowledgments**

We gratefully acknowledge L. Pichl (MPE) for assembling the tracker layers, F. Schrey (MPE) and K. Wölfl (MPE) for mechanical support, M. Kippen (LANL) for writing the GLECS code for GEANT3, and K.-H. Schenkl (Albedo) for electronics work.